# Universal model of the growth of III-V tetrahedral semiconductors on polar surfaces: pyramids and glue atoms. I. Clean surface properties

Pawel Strak, Stanisław Krukowski
Institute of High Pressure Physics, Polish Academy of Sciences, Sokołowska 29/37, 01-142 Warsaw, Poland

**Abstract**

Ab initio calculations were used to elucidate the properties of the polar surfaces of III-V tetrahedral semiconductors, i.e., metal- and nonmetal-terminated. The systems considered were clean wurtzite gallium nitride (i.e. $(0001)$ and $(000\bar{1})$ GaN surfaces), zinc blende gallium phosphide (i.e. $(111)$ and $(\bar{1}\bar{1}\bar{1})$ GaP), gallium arsenide (i.e. $(111)$ and $(\bar{1}\bar{1}\bar{1})$ GaAs), and $(111)$ and $(\bar{1}\bar{1}\bar{1})$ GaSb surfaces. The calculated data include the crystallographic structure and the electronic states of the termination atoms on both sides, i.e. their positions, energies, and occupations. All nonmetal-terminated surfaces exhibit no reconstruction, and their broken-bond states are located in the vicinity of the valence band maxima (VBM), pinning the Fermi level. The Ga-terminated surface exhibited $(4 \times 4)$ reconstruction related to the division of termination atoms into $sp^3$ and $sp^2$ hybridized in proportions of 6 to 10, based on charge redistribution according to the electron counting rule (ECR). Thus, the broken-bond state is separated into two sets, occupied and empty, with the Fermi level located between them in the vicinity of the conduction band minimum (CBM).

## I. Introduction

Despite drastic differences within the family of III-V compounds, the thermodynamic properties of a wide variety of semiconductors exhibit some common features [1,2]. These features are identified as the main factors leading to a specific growth mode that incorporates similarity of the atomic mechanisms of surface diffusion, sublimation by the creation of nonmetallic diatomic molecules and metallic atoms, and incorporation of diatomic radicals $(Me - N)$ at a kink in an atomic step during its motion. To exemplify these similarities, a series of gallium compounds, i.e. GaN, GaP, GaAs, and GaSb, will be discussed.

The common thermodynamic features of these systems are intrinsically connected to the bonding of nonmetal atoms into diatomic or tetra-atomic molecules [2-4]. Metal vapors are atomic. Primarily, the group III and V atoms differ by the number of electrons in the outer $s - p$ shells: group III atoms have three electrons that occupy the two $s$ states, and the one is located in $p$ state. Therefore, tetrahedral bonding involves promoting electrons from the lower energy $s$ states to common $sp^3$ orbitals, which is costly. This is standard in both metal and nonmetal bonding [5]. The only exceptions are nitrides, where the nitrogen orbitals are not hybridized. This hybridization is too costly because the difference between the energies of the nitrogen s and p states is extremely large. Therefore, nitrides are bonded differently: metal $sp^3$ orbitals create bonding overlaps with $p$ orbitals of nitrogen, and separate metal $d$ orbitals create bonding overlaps with nitrogen $s$ orbitals. The latter contributes to isotropic cohesion, and the former determine lattice symmetry [6]. Bonding by two $p$ electrons does not provide sufficient energy gain; thus, the group III atoms do not create diatomic molecules in the vapor phase. Consequently, their energy is relatively high, whereas the saturated vapor pressures above the condensed phase are low. Gallium compounds were selected because gallium is characterized by extremely low partial pressures in the vapor [7].

The group-V atoms have five electrons in the outer $s - p$ shells. Thus, their bonding to diatomic molecules involves five electrons. As a result, the molecular binding energies are high, and the saturated vapor pressures are correspondingly high. In this series, nitrogen vapor consists exclusively of diatomic molecules, that is, $N_2$. Phosphorus and arsenic form diatomic molecules: $P_2$ and $As_2$, and can also form tetra-atomic molecules with a tetrahedral structure: $P_4$ and $As_4$. In the case of antimony, the vapor predominantly consists of $Sb$ atoms [7].

The molecular energy is of primary importance both for understanding the bonding of nonmetal atoms and for verifying ab initio calculations. The dissociation energy of the nitrogen molecule ($N_2$) is extremely high, the highest of all diatomic molecules. The experimental value is $\Delta E_{exp}^{diss}(N_2) = 9.756\ eV$ [8,9]. For phosphorus, the experimental value of di-molecule ($P_2$) dissociation energy is $\Delta E_{exp}^{diss}(P_2) = 5.079\ eV$ [9]. For arsenic, the diatomic molecular dissociation energy is $\Delta E_{exp}^{diss}(As_2) = 4.042\ eV$ [10]. Finally, antimony is atomic in the vapor; thus, the atomic state is used as the reference state in this study.

At the metal-terminated surface, the process leading to the adsorption of nonmetallic species is complex. Currently, this phenomenon is only partially understood. The most detailed picture was created for the nitride metal-terminated surfaces: GaN (0001) and AlN (0001). These results are qualitatively different for low- and high-coverage regimes. This difference is expected to be present in the other cases. A more precise description is given in the model section, which follows the section discussing the calculation procedure. The Results section presents the main new data and is followed by a summary and conclusion at the end.

The long-term objective of this research is to create an atomic-level model of the growth of III-V semiconductors that will describe both growth and doping in these semiconductors. Among other factors, doping depends on the position of the Fermi level; therefore, it is necessary to include a description of the electronic properties of these surfaces. The present paper is the first in a series and is devoted to determining the crystallographic and electronic properties of polar surfaces. As shown, the two types of surfaces are drastically different. Nevertheless, surfaces of the same polarity show remarkable similarities across the class, which is the basis of the project aimed at developing a unified model of growth and doping applicable to all tetrahedral III-V semiconductors.

The proposed approach is devoted exclusively to developing the vapor growth model. It is therefore assumed that the surfaces are at high coverages of both species. Therefore, the thermodynamic properties of the system leading to the loss (evaporation) of matter will not be considered because these are typical of a high-vacuum environment [11]. Gallium is not easily volatile; thus, the Ga-terminated stoichiometric surface is possible for the GaN(0001)A (Ga-terminated) surface [12]. The Ga-terminated surface will be shown to have charge-controlled weak reconstruction in accordance with the Peierls mechanism. On the opposite nitrogen-terminated surface, nitrogen loss is not rapid; thus, the stoichiometric surface may be attained [12]. The N-terminated surface has no symmetry change. Nevertheless, gallium or nitrogen atoms may be lost from Ga- and N-terminated polar GaN surfaces. In the following, we illustrate an adsorption-related stabilization mechanism allowing us to consider reconstruction-free surfaces as the basis for the development of the growth model of gallium nitride.

In the case of gallium phosphide, a richer variety of reconstructions is observed than in GaN. The P-terminated GaP(111)B surface shows two different reconstruction units [13]. Recently, extensive investigation of the surface reconstructions of the GaP(111)B surface established a phase diagram comprising $(2\ \times 2)$, $(1\ \times 1)$ $\left(\sqrt{19}\ \times\ \sqrt{19}\right)$ phases as a function of $PH_3$ flow and the temperature [14]. The opposite Ga-terminated GaP(111)A surface has not been extensively investigated.

In the case of GaAs, the surface chemical state is controlled by Ga or As evaporation, which leads to surface reconstruction associated with Ga vacancies or adsorbed As-trimers at GaAs$\{111\}$ A and B surfaces, respectively [15]. Depending on the As pressure, a number of reconstructions are possible on the GaAs$\{111\}$B surface, such as $\left(\sqrt{19}\ \times \sqrt{19}\right)$, $(1\ \times 1)$ and $(2\ \times 2)$ symmetries [15,16].

Finally, the gallium antimonide metal-terminated surfaces exhibit the most complex reconstruction patterns because of the relatively narrow band gap and weak bonding, which entail relatively high mobility of surface atoms. Owing to the large atomic radius of Sb, the creation of Sb dimers, trimers, and tetramers is favored. Accordingly, this mechanism is responsible for the Ga-terminated $GaSb(111)A$ surface $(12\ \times 1)$ or $(2\ \times 5)$, and $(1\ \times\ 5)$ reconstructions [17]. Similarly, the Sb-terminated $GaSb(111)B$ surface undergoes $(8\ \times\ 1)$ reconstructions [17].

In this paper, the structural and electronic properties of the metal- and nonmetal-terminated stoichiometric surfaces are investigated using ab initio calculations. These surfaces are considered under the assumption that the topmost atoms are stable, i.e. no surface vacancies are created. In subsequent papers (II and III) we will investigate the adsorption of the complementary species of the compound on metal- and nonmetal-terminated polar surfaces. It will be shown that the adsorption stabilizes the polar surfaces and prevents their destruction. Therefore, such covered surfaces will be used in the formulation and analysis of the vapor-growth-related processes.

## II. The calculation procedure

Ab initio density functional theory (DFT) calculations were used for a large number of simulations devoted to semiconductor crystal surfaces representing the III-V family. The crystals considered were gallium-based compounds: GaN, GaP, GaAs, and GaSb. For these extensive calculations, we selected the Spanish Initiative for Electronic Simulations with Thousands of Atoms (SIESTA) *ab initio* shareware package [19]. The main task is to solve the Kohn-Sham equations to determine the electronic properties of finite systems with periodic boundary conditions (PBC) imposed. The solutions are eigenvalues and the corresponding eigenfunctions, which are expressed as a series of specific finite-range numerical orbitals centered on the atoms present in the system [20,21]. The orbitals are the products of angular and radial functions, the latter having finite cutoff radii. The angular dependence is expressed by standard spherical harmonics $Y_{l,m}(\theta,\varphi)$ for $l \leq 2$, i.e., s, p, and d orbitals. The s and p orbitals of gallium and nonmetals, i.e., nitrogen, phosphorus, arsenic, and antimony atoms, are represented by triple zeta functions. In the case of gallium, the internal d shell electrons are incorporated into the valence electrons, which are represented by single zeta functions. The nonmetal atoms have d electrons included in the atomic cores. The functional basis is greatly reduced by the application of Troullier-Martins pseudopotentials [22,23].

Integration in reciprocal k-space is approximated by summation over the Monkhorst-Pack grid, which varies among the simulated systems [24]. The method is an application of the 19th-century quadratures proposed by Johann Carl Friedrich Gauss [25]. The GGA-PBE (PBEJsJrLO) density functional was parameterized using the β, μ, and κ values set by the jellium surface (Js), jellium response (Jr), and Lieb-Oxford bound (LO) criteria [26,27].

The electrostatic-potential problem, tackled using the linear Poisson equation, is solved using the fast Fourier transform (FFT), which is superior to any other method. This implies periodic boundary conditions that create coupling across the vacuum space owing to the surface-slab-dipole-induced potential difference [28]. The periodicity induces artificial interaction between slab copies that must be eliminated either by introducing a compensating dipole in the vacuum space [28,29] or by the additional Laplace solution contribution to remove the field in the space between slab copies [30]. Both methods are applicable because they are essentially equivalent.

Under normal conditions, GaN has a hexagonal wurtzite structure with the following experimentally determined lattice parameters: $a_{GaN}^{exp} = 3.189$ Å, $c_{GaN}^{exp} = 5.186$Å [31]. The calculated GaN lattice parameters were $a_{GaN}^{DFT} = 3.216$ Å and $c_{GaN}^{DFT} = 5.233$ Å in good agreement with the experimental values.

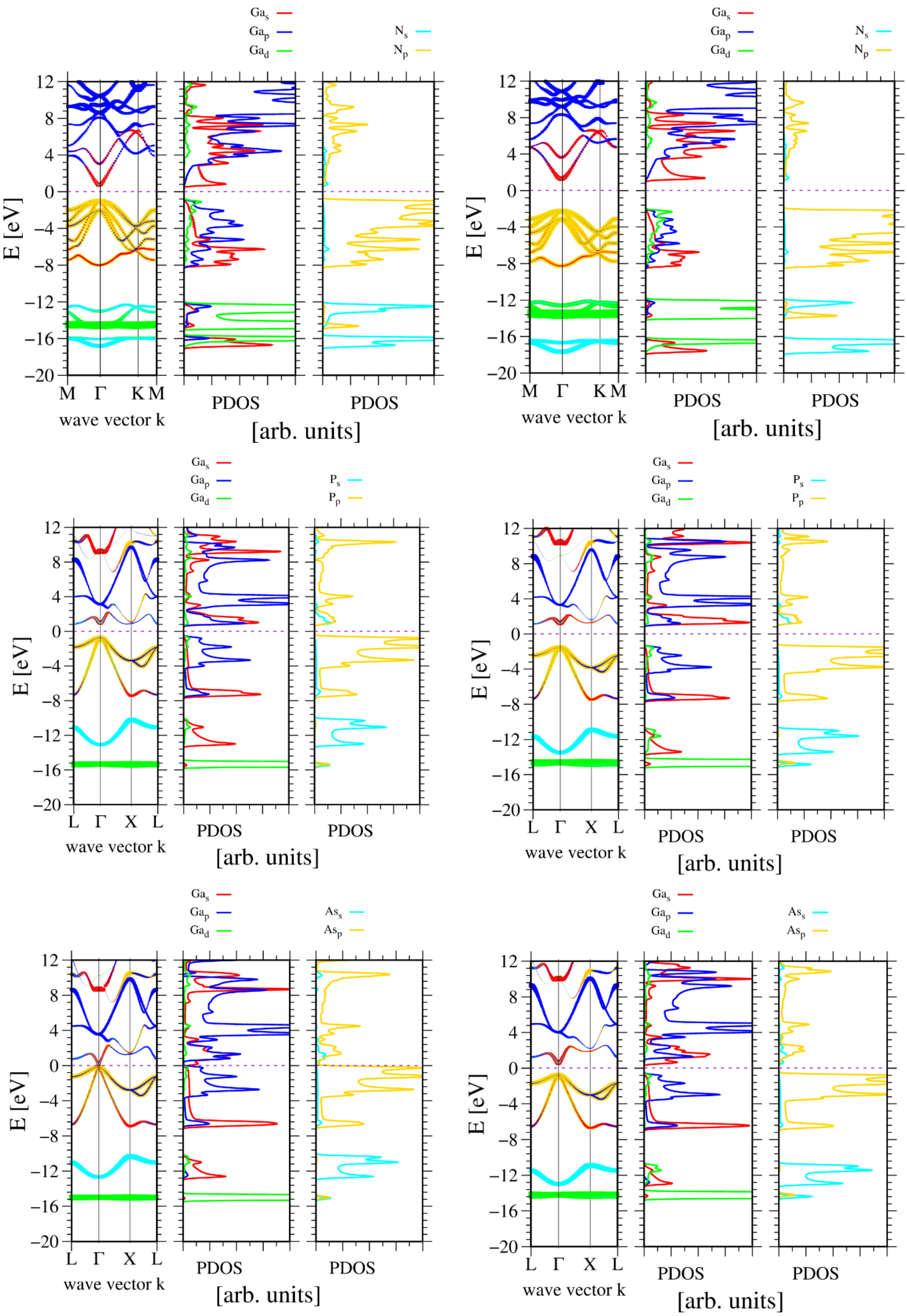
$Ga_s$
$Ga_p$
$Ga_d$
$N_s$
$N_p$
$P_s$
$P_p$
$As_s$
$As_p$
E [eV]
12
8
4
0
−4
−8
−12
−16
−20
M Γ K M
L Γ X L
wave vector k
PDOS
[arb. units]

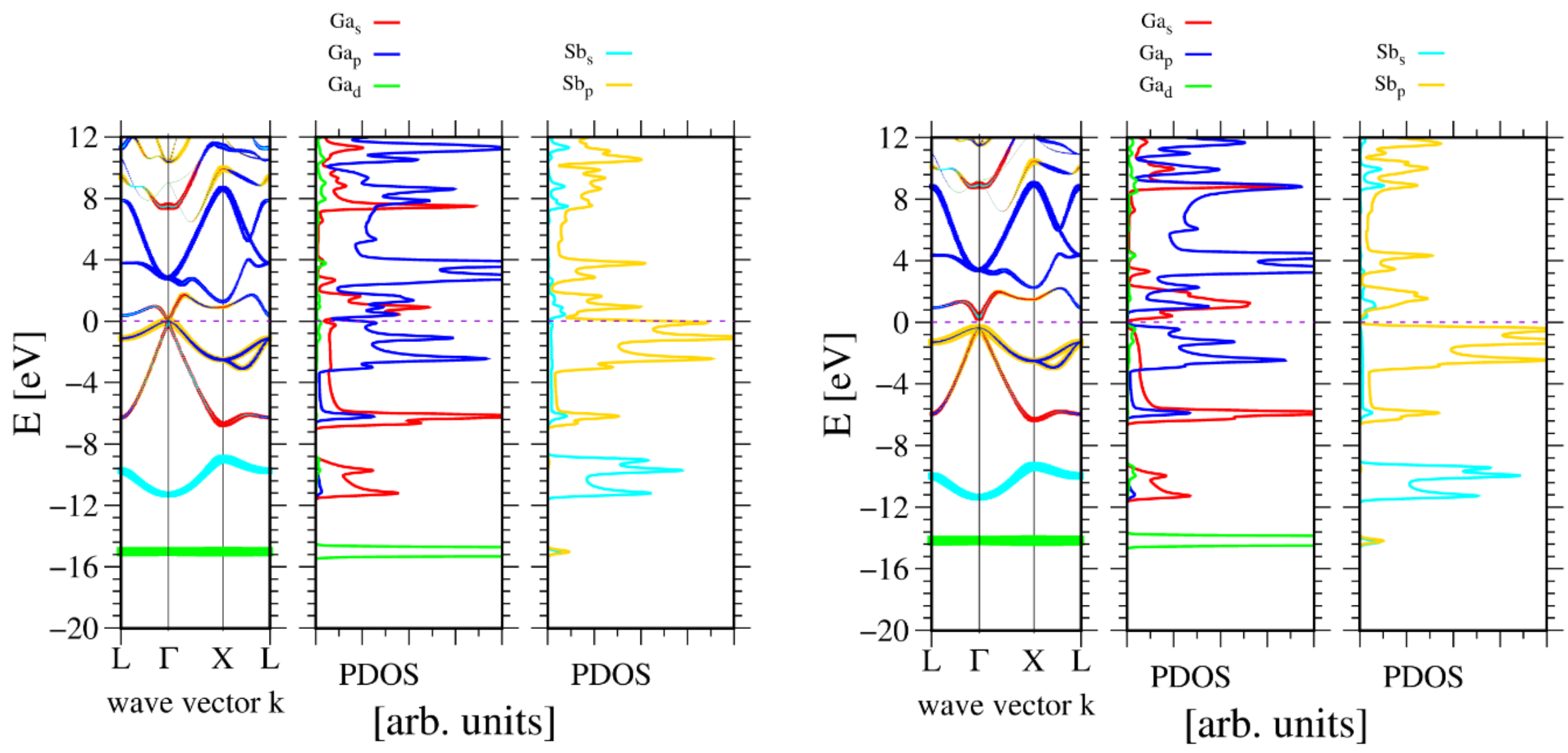

*Fig. 1. Band diagrams and partial densities of states (PDOS) projected on metal (Ga) and nonmetal atomic states obtained directly using PBEJsJrLO exchange-correlation (left column) and Ferreira correction (right) for the III-V compounds: (a) GaN, (b) GaP, (c) GaAs, (d) GaSb.*

The band correction scheme of Ferreira et al., known as GGA-1/2, was used to obtain improved band-gap energies, effective masses, and band structures [32,33]. The electronic properties were obtained using a modified Ferreira scheme, for which the positions of atoms and the periodic cell were first obtained using the PBEJsJrLO exchange-correlation functional. The band diagrams obtained using these two approximations for the four III-V compounds, GaN, GaP, GaAs, and GaSb, are presented in Fig. 1.

The *ab initio* bandgap using PBEJsJrLO functional was $E_g^{DFT}(GaN) = 1.77\ eV$ while the corrected (GGA-1/2) was $E_g^{DFT}(GaN) = 3.51\ eV$. Thus, the Ferreira correction is essential to obtain good agreement with the low-temperature experimental data: $E_g^{exp}(GaN) = 3.47\ eV$ [1,35,36]. These energy band data showing the bonding scheme are generally consistent with the recent results obtained for nitrides, indicating the existence of two subbands in the valence band (VB): an upper subband created by the overlap of hybridized $Ga4sp^3$ states and nitrogen $N2p$ states, and a lower subband composed of $Ga3d$ and $N2p$ states [1,6]. This division into VB higher and lower subbands will be intensively used in the following: the upper subband states and their counterparts in the case of adatoms create bonding/antibonding states that are occupied or empty, depending on the position of the Fermi level, thus they will be explicitly considered. Thus in the case of gallium these are $Ga4s$, $Ga4p$ and also $Ga4sp^3$ hybridized states. In the case of nitrogen these will be nitrogen $N2p$ states. The lower GaN VB subband states and also the adatom $Ga3d$ and $N2s$ states will be tacitly assumed to be fully occupied, independent both in the crystal bulk and on the surface. Nevertheless, these states contribute to bonding both in the bulk and on the surface stabilizing the system. Because they are always occupied, their explicit consideration is not essential and therefore they will not be shown in the following. The only exception will be antibonding which be additional states to be occupied and also higher in energy, so they will be considered.

Gallium phosphide has a cubic zinc-blende structure with the measured lattice parameter $a_{GaP}^{exp} = 5.4495$ Å [1,2]. The *ab initio* calculations gave $a_{GaP}^{DFT} = 5.49$ Å which is in reasonable agreement with the experiment. Experimentally, GaP was determined to have an indirect bandgap $E_g(GaP) = 2.24\ eV$[1]. The *ab initio* calculations using the PBEJsJrLO functional yielded an indirect gap $E_g^{DFT}(GaP) = 1.66\ eV$, whereas the GGA-1/2 Ferreira scheme resulted in a direct gap $E_g^{DFT}(GaP) = 2.74\ eV$ . As in GaN, the VB in GaP is divided into two separate subbands: the upper composed of $Ga4s$, $Ga4p$ and $P3p$ states and the lower due to $Ga3d$ and $P3s$ states. Accordingly, the latter are always occupied so they will not be considered explicitly in the analysis of surface processes. The only exception will be made in case of antibonding states that could be potentially occupied.

Gallium arsenide also has a cubic zinc-blende lattice with the parameter $a_{GaAs}^{exp} = 5.65315$ Å [1,2]. The *ab initio* test calculations resulted in lattice parameter: $a_{GaAs}^{DFT} = 5.71$ Å, i.e., in satisfactory agreement with experiment. The band gap of GaAs is direct at $E_g^{exp}(GaAs) = 1.424\ eV$ at $T = 300\ K$[1,2]. The *ab initio* calculations using the PBEJsJrLO functional resulted in a direct narrow gap $E_g^{DFT}(GaAs) = 0.28\ eV$. The Ferreira scheme increased the gap to $E_{GaP}^{DFT} = 1.33$ eV. This result is in reasonable agreement with the experimental data. In the case of GaAs the pattern observed in the two preceding cases is again observed: VB is divided into two separate subbands: the upper composed of $Ga4s$, $Ga4p$ and $As4p$ states and the lower related to $Ga3d$ and $As4s$ states. By analogy with the preceding cases the latter states will be routinely omitted.

Finally, gallium antimonide under standard conditions has a stable cubic zinc-blende structure with a lattice parameter: $a_{GaSb}^{exp} = 6.0959$ Å at $T = 300\ K$[1]. The *ab initio* calculations at 0 K gave similar results: $a_{GaSb}^{DFT} = 6.18$ Å. The band gap of GaSb is narrow and direct $E_g^{exp}(GaSb) = 0.812\ eV$ at $T = 300\ K$[1,2]. The ab initio calculations based on the PBEJsJrLO functional yield a closed gap. The application of the GGA-1/2 correction (Ferreira scheme) opened the gap to yield a direct band gap $E_g^{DFT}(GaSb) = \ 0.68\ eV$. In summary, the Ferreira correction to the ab initio calculations provided satisfactory agreement of the structural and energetic parameters of all three zinc-blende III-V compounds suitable for the analysis of the most important properties of the polar surfaces. GaSb follows again the trends of separation into subbands: the upper due to overlap of $Ga4s$, $Ga4p$ and $Sb5p$ states and the lower related to $Ga3d$ and $Sb5p$ states. The latter will not be considered in most of the investigations in the following.

It is also important to consider the differences between the energies of the nonmetal $s$ and $p$ states. For GaN, the weighted energy difference was approximately $13\ eV$ whereas for GaP, GaAs, and GaSb, these differences were smaller (approximately $8\ eV$). Thus, the energy gain due to the increase in the proportion of s states is much larger for gallium nitride than for the other compounds.

In these calculations, the self-consistent-field (SCF) loop was terminated when the difference for all elements of the density matrix in two consecutive iterations was smaller than $10^{-4}$. A real-space grid is used for the calculation of the multicenter overlap integrals that are controlled by the energy cut-off. The cutoff value is 410 Ry, which is equivalent to a grid spacing in real space of 0.08 Å.

For calculations of surface reactions with vapor species, the molecular energy is of primary importance, including both the experimental value and the theoretical estimate. Thus, the dissociation energy of the nitrogen diatomic molecule, obtained from ab initio calculations, is $\Delta E_{DFT}^{diss}(N_2) = 9.86\ eV$. This is in very good agreement with the experimental value $\Delta E_{exp}^{diss}(N_2) = 9.756\ eV$ [8,9]. For the phosphorus diatomic molecule, the ab initio approach gave $\Delta E_{DFT}^{diss}(P_2) = 5.15\ eV$, which is again in good agreement with experiment $\Delta E_{exp}^{diss}(P_2) = 5.079\ eV$ [9]. Finally, the theoretical value for the arsenic diatomic molecule is $\Delta E_{DFT}^{diss}(As_2) = 4.13\ eV$, which is in agreement with the experimental data giving $\Delta E_{exp}^{diss}(As_2) = 4.042\ eV$ [10]. For antimony, the atomic state was used as the reference throughout these publications. For the nitrogen tetratomic molecule is not identified [36,37]. For phosphorus and arsenic, the tetratomic molecules are stable. Their atomization energies are $\Delta E_{DFT}^{diss}(P_4) = 12.40\ eV$ and $\Delta E_{DFT}^{diss}(As_4) = 10.25\ eV$, respectively [38,39]. These tetratomic molecules will not be used as reference throughout the publications in this series.

The Born-Oppenheimer approximation was enforced in the procedure for determining the energy barriers. The method is based on the nudged elastic band (NEB) method to optimize the path that has both endpoints fixed in the optimization [40-42]. The minimum-energy pathway (MEP) is characterized by a saddle point that establishes the energy barrier. The NEB module was linked to SIESTA to determine the minimum barrier energy and path of the species. Atomic relaxation was terminated when the forces acting on the atoms were below 0.005 eV/Å.

## III. Model and results: equilibrium surface structure

### 1. Nonmetal (N, P, As, and Sb) side

The physical properties of nonmetal (group-V-atom)-terminated clean semiconductor surfaces, such as $GaN(000\bar{1})$, $GaP(11\bar{1})$, $GaAs(11\bar{1})$ and $GaSb(11\bar{1})$ are similar in terms of crystallographic structure and electronic properties. As shown in Fig. 2, the results of ab initio slab simulations indicate the absence of any reconstruction on all surfaces. The opposite surfaces (i.e. Ga-terminated surfaces) of all nonmetal-terminated slabs have their bonds saturated by fractionally charged hydrogen $(\mathrm{Z} = 5/4)$ pseudoatoms.

(a)

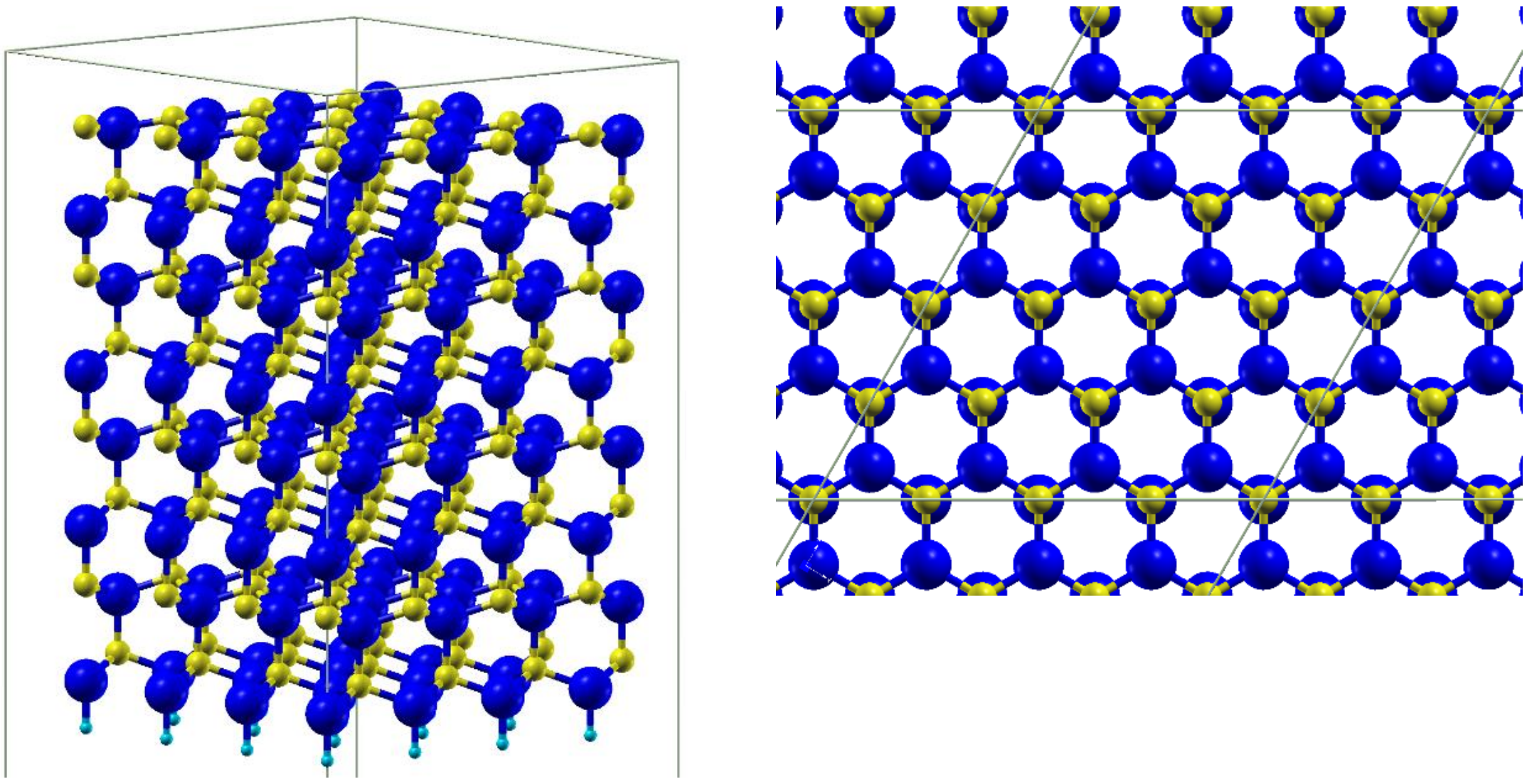

(b)

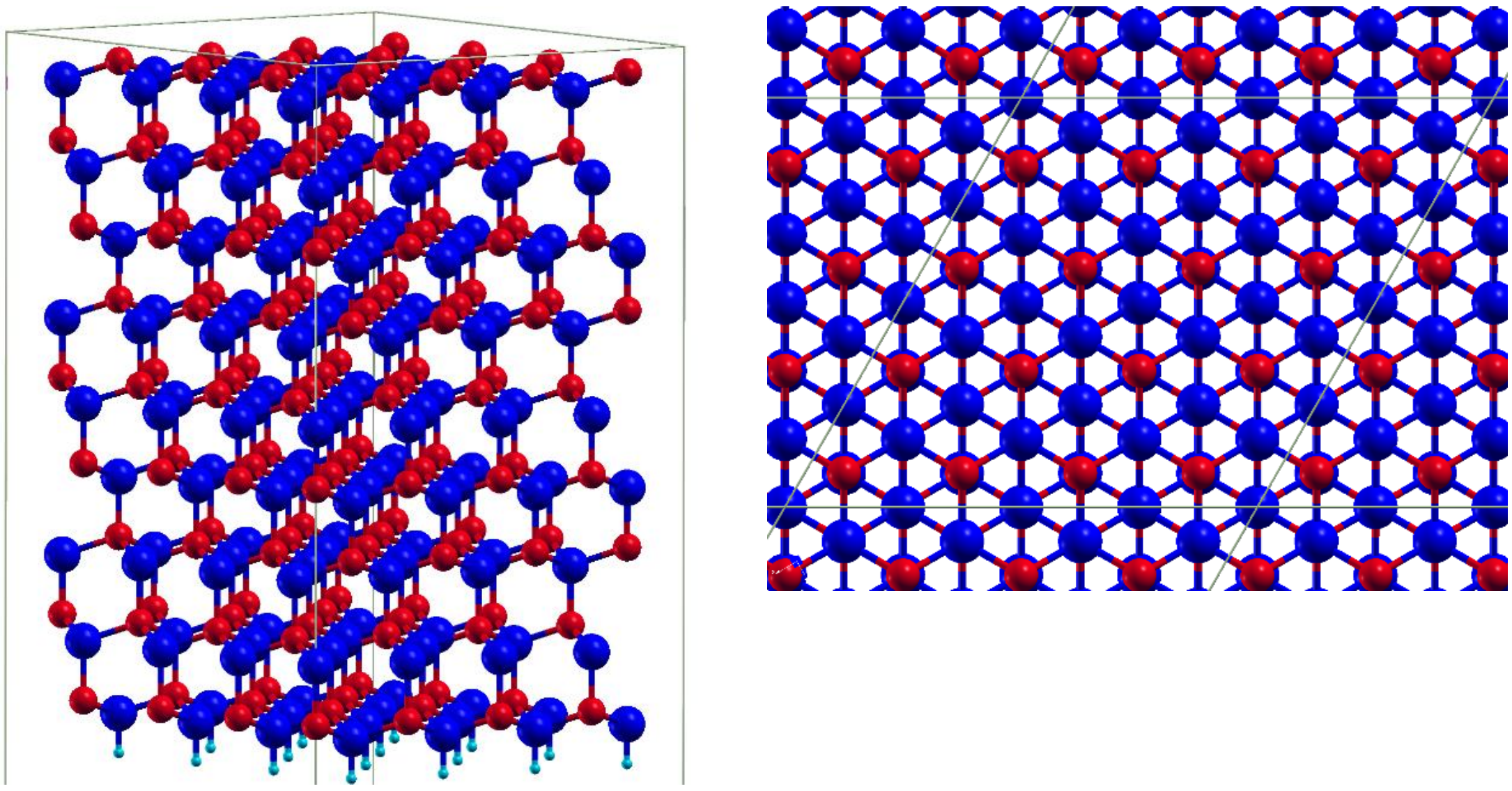

(c)

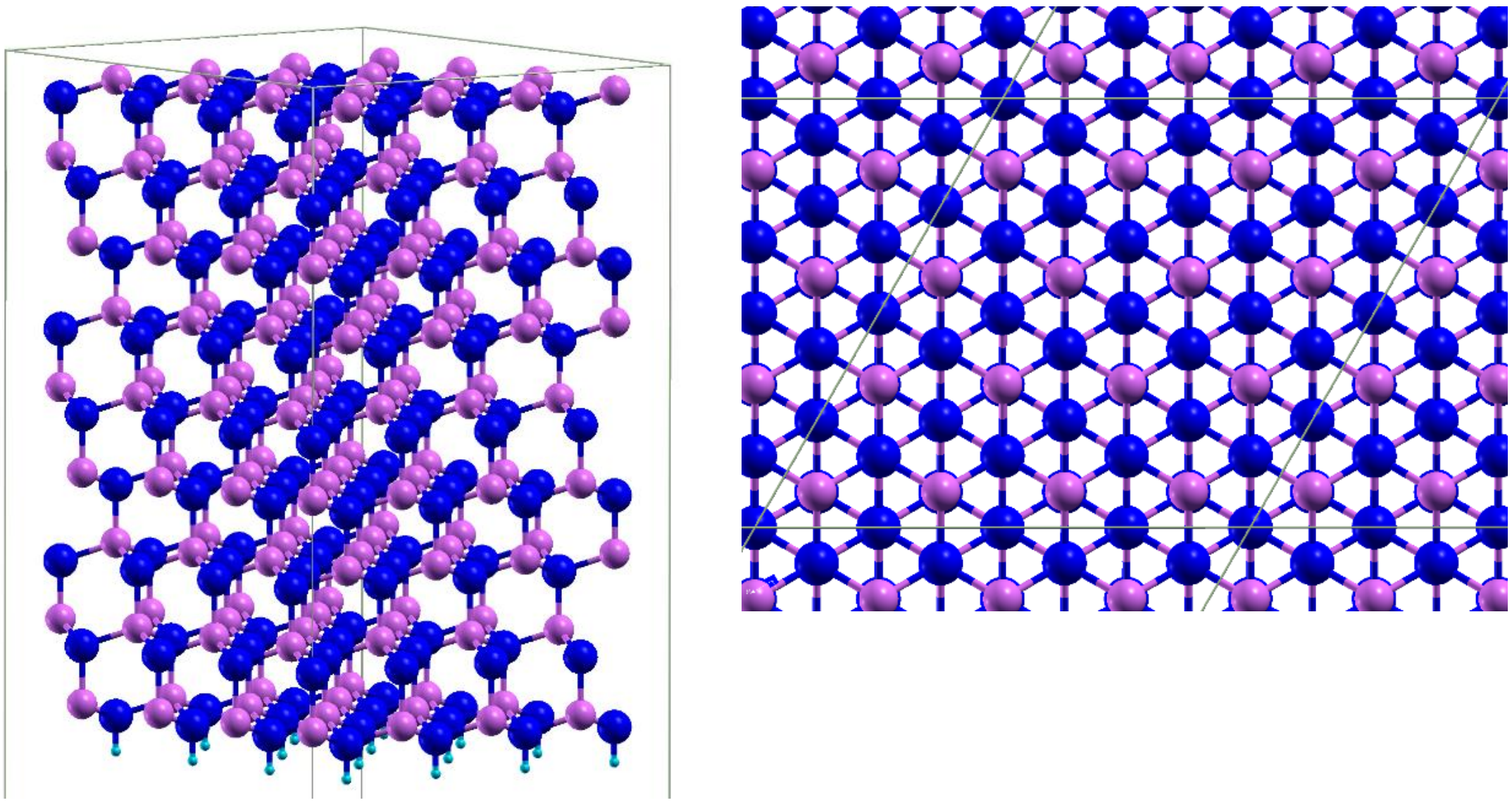

(d)

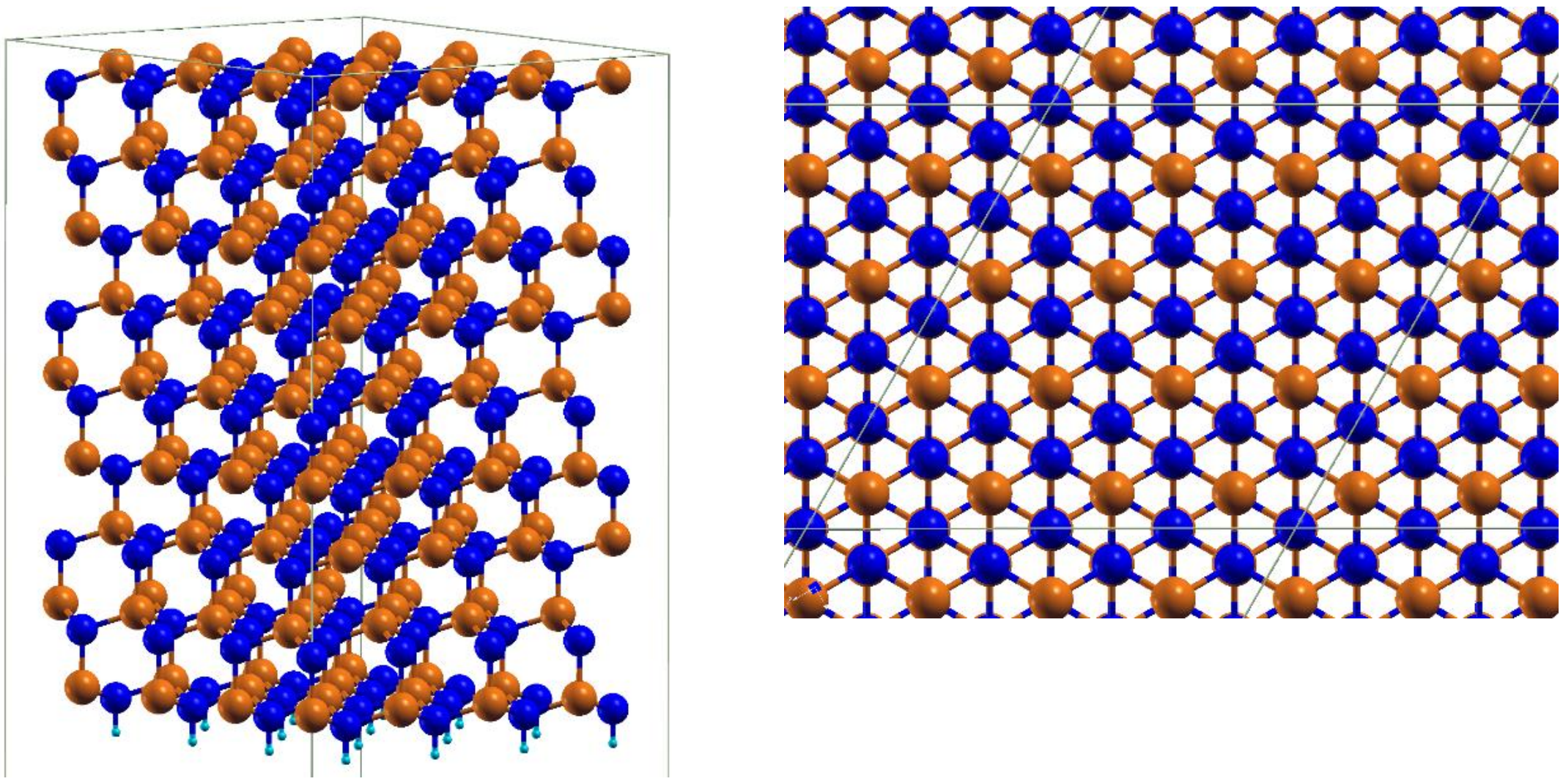

*Fig. 2. Crystallographic structure of energetically stable nonmetal-terminated surfaces of III-V semiconductors obtained using* $(4 \times 4)$ *slabs: (a) 8 double atomic layer (DALs)* $GaN(000\bar{1})$*, (b) 9 DALs* $GaP(\bar{1}\bar{1}\bar{1})$*, (c) 9 DALs* $GaAs(\bar{1}\bar{1}\bar{1})$*, (d)* $GaSb(\bar{1}\bar{1}\bar{1})$*. Gallium, nitrogen, phosphorus, arsenic, and antimony atoms, are represented by blue, yellow, red, lavender and orange balls, respectively. In the following papers of this series the opposite surfaces of all slabs representing the nonmetal-terminated surfaces are saturated by fractionally charged hydrogen* $(Z = 5/4)$ *pseudoatoms, represented by small sky-blue balls.*

These results indicate that the energy reduction due to relaxation is limited, and is insufficient to shift the atoms from the lattice registry. Nevertheless, the topmost nitrogen atoms are located below the plane of the gallium atoms. In other cases, the nonmetal atoms were located above the Ga atoms.

Consistent with the structural properties, the electronic properties of all four nonmetal-terminated surfaces are remarkably similar. The nonmetal broken-bond state is located slightly above the valence band maximum and is independent of the bandgap. The only difference is that, in the case of GaN, this is the pure N $4p_z$ state in accordance with the lattice symmetry, in which N atoms are located in the plane of the Ga atoms, i.e., bonded through a $sp^2$ hybridized state. For the other zinc-blende compounds, these are fully hybridized $sp^3$ states. Naturally, these states are partially occupied, thereby pinning the Fermi level close to the VBM, which is typical for p-type semiconductors.

(a)

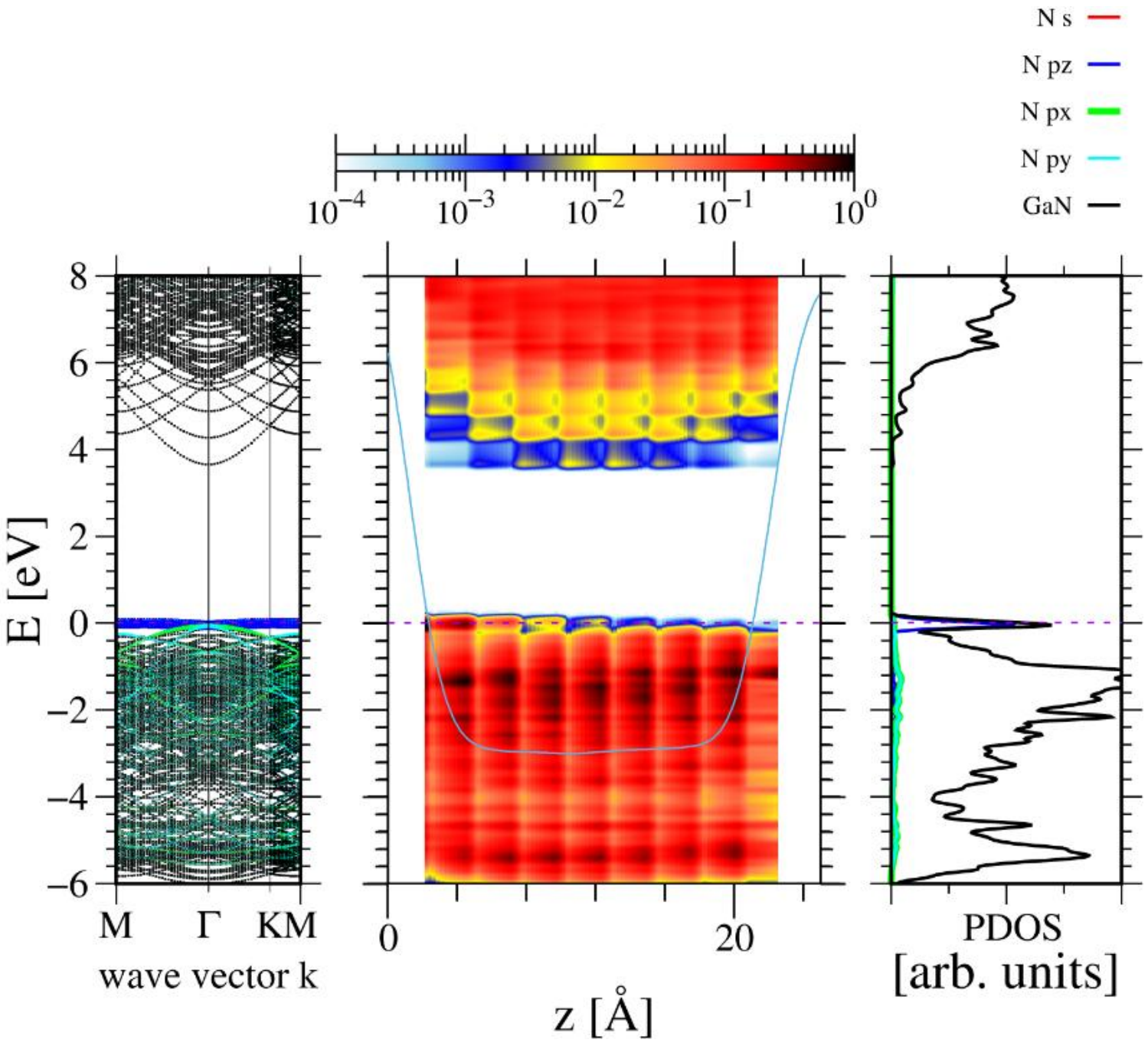

(b)

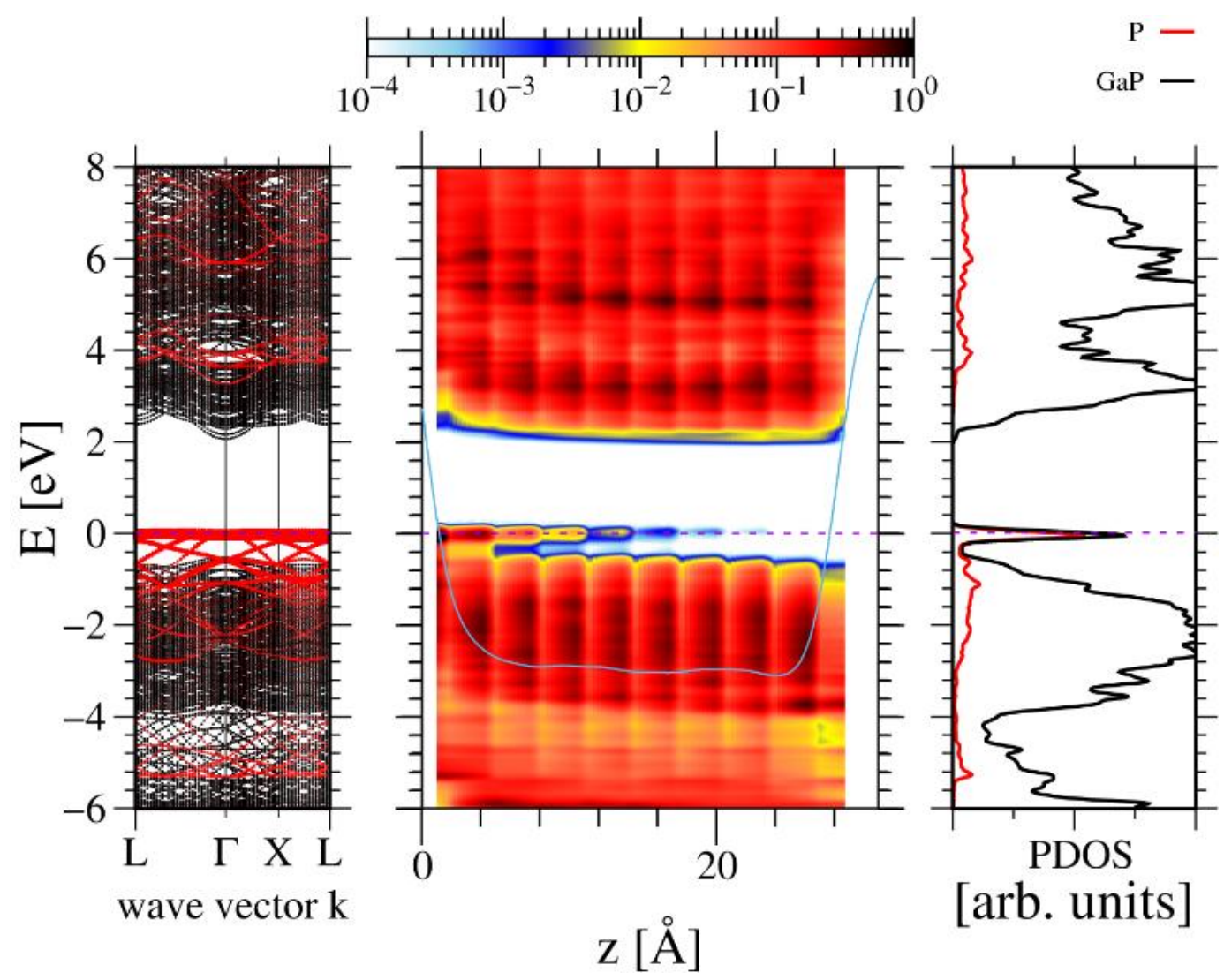
P
GaP
E [eV]
wave vector k
z [Å]
PDOS
[arb. units]

(c)

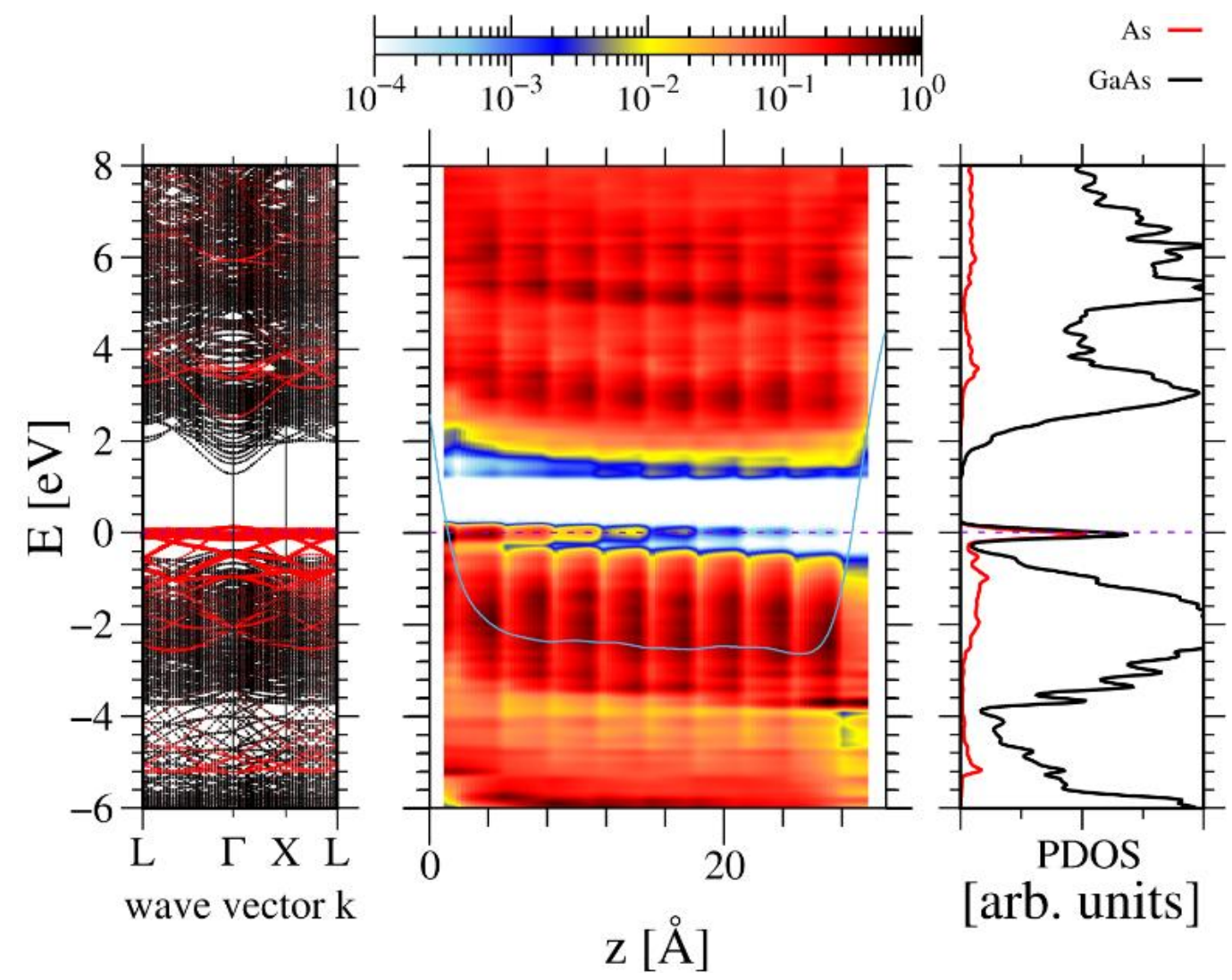
As
GaAs
E [eV]
wave vector k
z [Å]
PDOS
[arb. units]

(d)

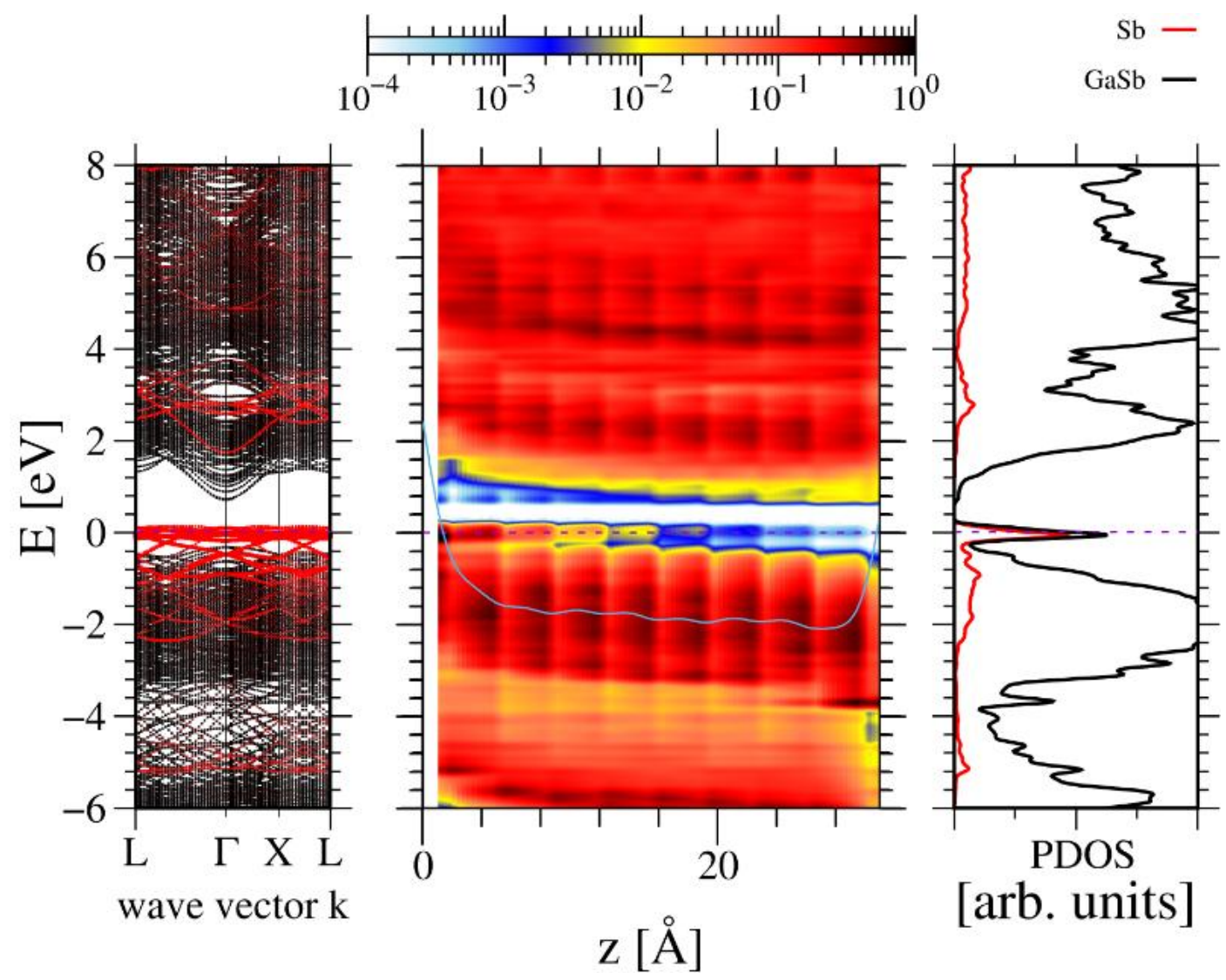
Sb
GaSb
E [eV]
wave vector k
z [Å]
PDOS
[arb. units]

*Fig. 3. Electronic properties of nonmetal-terminated (0001) Ga-V surface represented by 8 and 9 Ga-V DALs slabs for wurtzite and zinc blende structures respectively: band diagrams in momentum (left panel) and position space oriented along c-axis (middle panel). PDOS of nonmetal surface atoms are shown in the right panel.*

The nonmetal-terminated surfaces do not exhibit separation into regions with $sp^2$ and $sp^3$ reconstructions. For GaN nitrogen, the $p_z$ states have energies equal to the valence band states; therefore, they cannot be empty. Thus, the total evacuation of the states is not possible, and all states are fractionally occupied and all nitrogen surface atoms remain in $sp^2$ hybridization. On the other surfaces, the effective equality of the energies of the valence band maximum states and the nonmetal $sp^3$ broken-bond states of the GaP, GaAs, and GaSb surfaces leads to their partial occupation. Therefore, these states could not be fully occupied, as expected from the charge shift from $sp^3$ to $sp^2$. Thus, they remain in $sp^3$ hybridized state. The opposite side of the slab, i.e., the metal-terminated surface, exhibits drastically different properties, both in terms of crystallographic structure and electronic properties, which are related to the charge-controlled $sp^2$ and $sp^3$ reconstruction.

Table 1. Structural and electronic properties of nonmetal (NM)-terminated polar stoichiometric surfaces

| Surface | Structure | Fermi level – VBM difference |
|---|---|---|
| $GaN(000\bar{1})$ | $sp^2$ - uniform | ~ 0.0 |
| $GaP(\bar{1}\bar{1}\bar{1})$ | $sp^3$ - uniform | 0.5 eV |
| $GaAs(\bar{1}\bar{1}\bar{1})$ | $sp^3$ – uniform | 0.4 eV |
| $GaSb(\bar{1}\bar{1}\bar{1})$ | $sp^3$ - uniform | 0.3 eV |

As shown, these properties are relatively uniform across all considered compounds.

## 2. Metal (gallium) side

The structural and electronic properties of the gallium-terminated surfaces of all compounds, i.e., GaN, GaP, GaAs, and GaSb, are drastically different from those of the nonmetal side. In fact, this is related to metal states having a much higher energy. Therefore, the broken-bond states are located high in the band gap, close to the conduction band minimum (CBM). Nevertheless, these states were similar in all members of the investigated group.

The atomistic structure and electronic properties of metal-terminated GaN(0001) surfaces have been extensively investigated. The reported results changed considerably over time, primarily associated with the size of the simulated systems; hence, we only describe the final version [45]. The structure of the slab representing the Ga-terminated $(0001)\ GaN$ surface is illustrated in Fig. 4.

(a) (b)

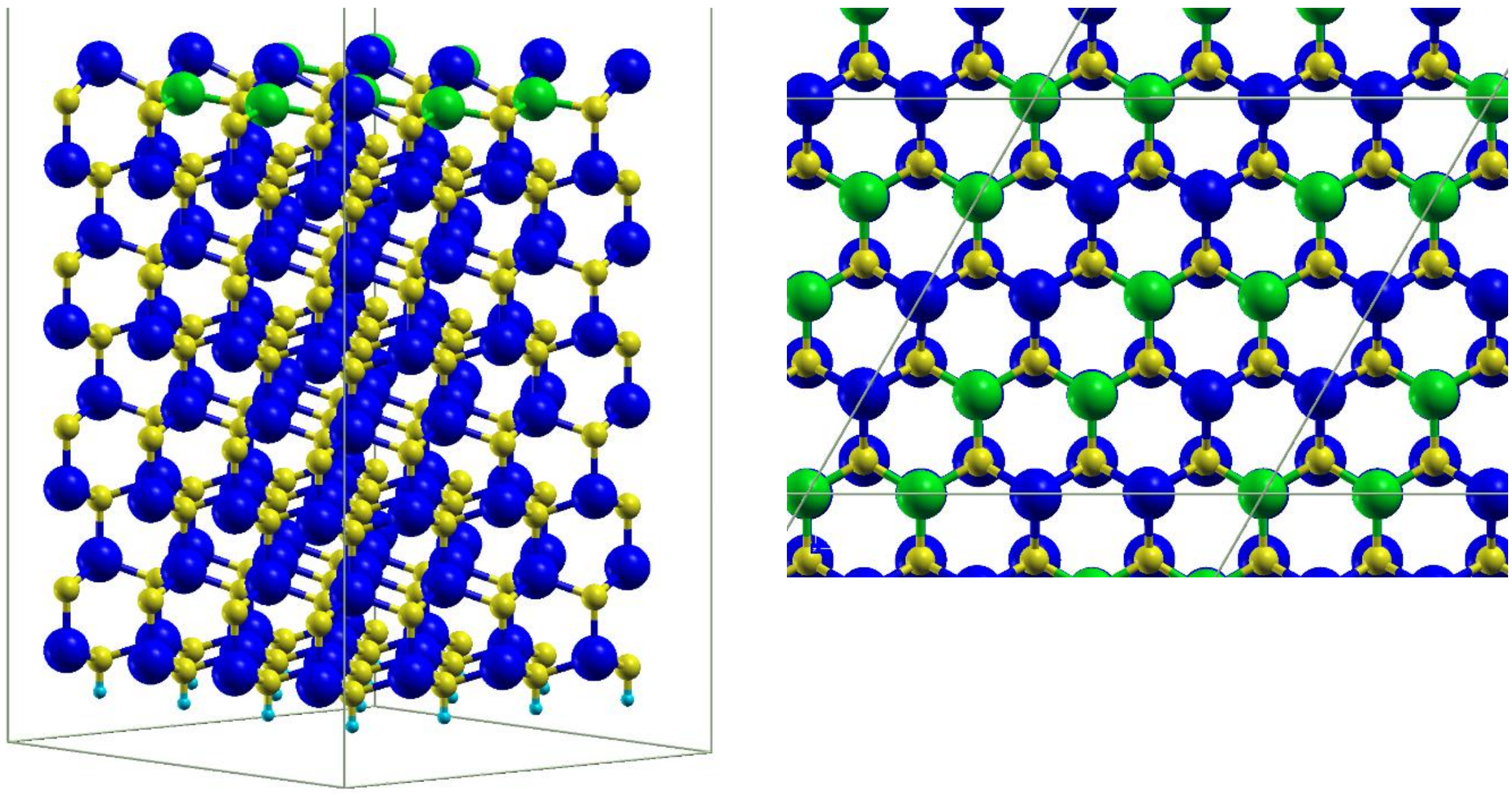

*Fig. 4. Metal-terminated GaN (0001) surface: the atoms in $(4 \times 4)$ 8 Ga-N double atomic layers (8 DAL) slab, (a) side view, (b) top view. Gallium, nitrogen and pseudo-hydrogen atoms are denoted by blue, yellow, and sky-blue balls. The topmost surface atoms in $sp^2$ hybridization are denoted by green balls. In the following papers of this series the opposite surfaces of all slabs representing the Ga-terminated surfaces are saturated by fractionally charged hydrogen $(Z = 3/4)$ pseudoatoms, represented by small sky-blue balls.*

The broken-bond states of the topmost Ga atoms (Ga-BB states) can be divided into two categories due to separation of topmost Ga surface atoms into $sp^3$ and $sp^2$ hybridization states.

As shown in Fig. 5, the latter have $Ga4p_z$ state empty due to the Peierls transition that opens the gap within the Ga broken-bond-state band that is located close to the conduction band minimum (CBM). This is because the fraction of the electrons is transferred to the bulk conduction band as the $sp^3$ hybridized surface state is located partially above the Fermi level. Thus, the surface was identified as metallic because Ga-BB states have considerable dispersion [45].

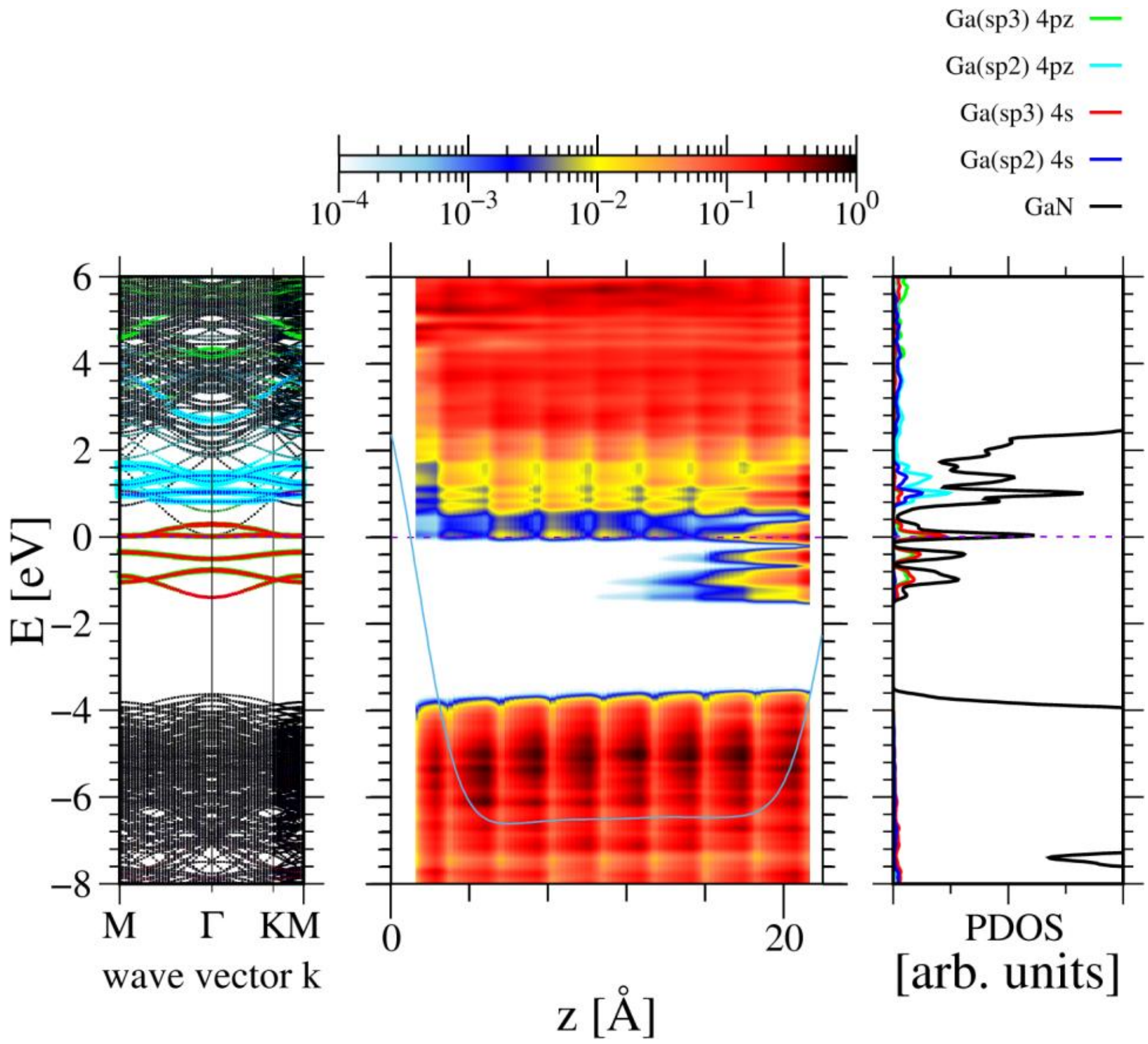

*Fig. 5. Electronic properties of the metal- terminated GaN (0001) surface, represented by 8 Ga-N DAL slab: band diagrams in momentum (left panel) and position space oriented along c-axis (middle panel). PDOS of gallium surface atoms are presented in the right panel.*

According to recent investigations, the Ga-terminated GaN surface undergoes $(4 \times 4)$ reconstruction [39,40]. The periodicity of the reconstruction may be explained using the electron counting rule (ECR), originally proposed by Chadi [46]. According to the ECR argument, each Ga surface atom contributes $^3/_4$ electrons. These electrons may occupy fraction $\alpha$ of the Ga atom's spin pair of $sp^3$ hybridized states. The remaining $1 - \alpha$ Ga atoms have $p_z$ states empty. Thus, this rule yields $2\,\alpha = {}^3/_4$, i.e. $\alpha = {}^3/_8$. Thus, the reconstruction pattern that includes the integer number of both hybridization states of the Ga atoms is 16. Of the 10 atoms remaining in $sp^2$ hybridization, the remaining 6 are in the standard $sp^3$ hybridization [43,44]. This is in approximate accordance with the structure presented in Fig. 4.

The electronic properties of the GaP(111) surface had not previously been determined by *ab initio* calculations. Such calculations were performed using $(4 \times 4)$ slab containing eight double Ga-P atomic layers, with a fractionally charged hydrogen $(Z = 3/4)$ pseudoatom termination at the opposite surface. The resulting surface structure again indicated $sp^2$ - $sp^3$ reconstructions.

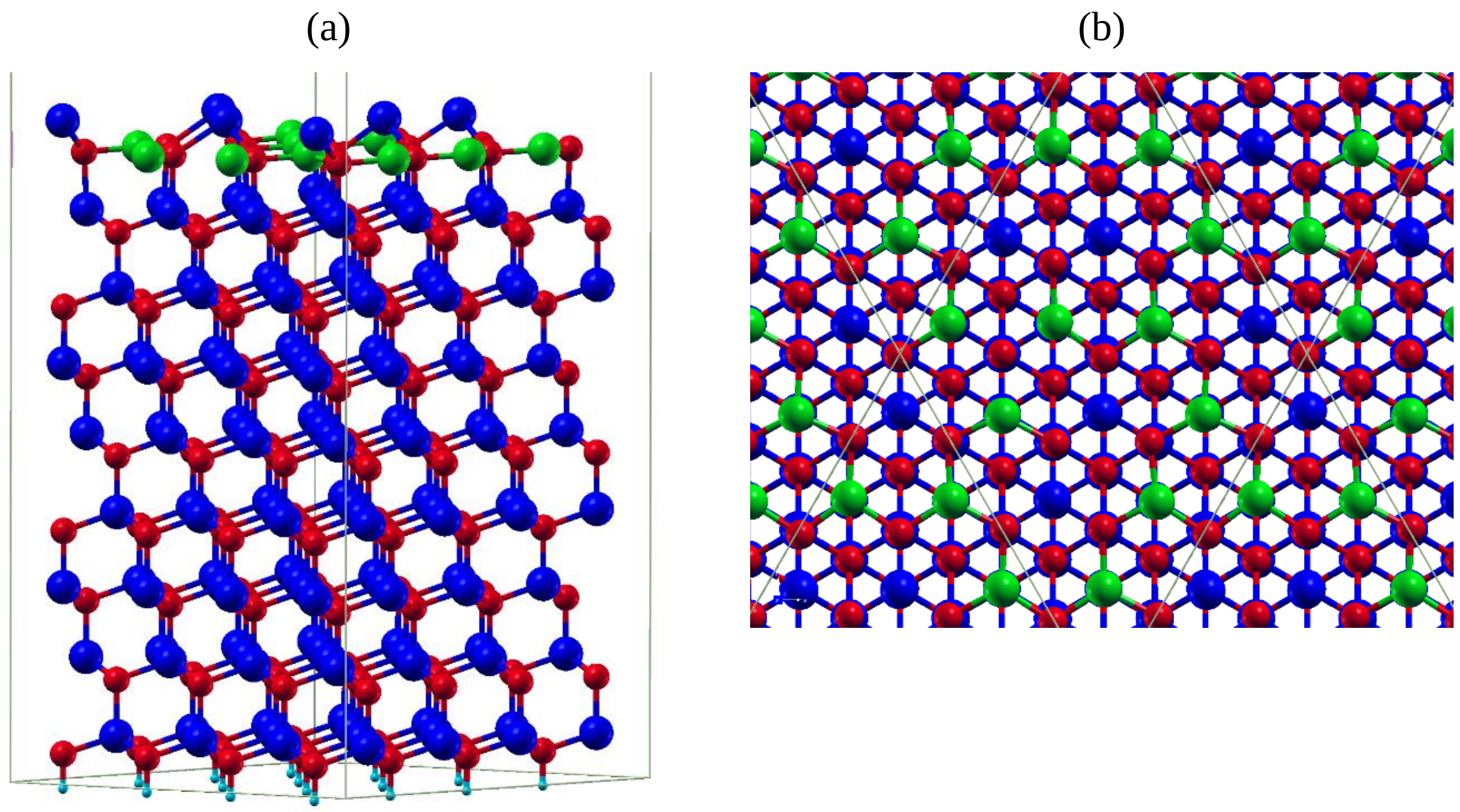

*Fig. 6. Metal-terminated GaP (111) surface: the atoms in* $(4 \times 4)$ *9 Ga-P DALs slab, (a) side view, (b) top view. Gallium, phosphorus and pseudo-hydrogen atoms are denoted by blue, red and sky-blue balls. The topmost surface atoms in* $sp^2$ *hybridization are denoted by green balls.*

The electronic states of $(111)\,GaP$ surface are shown in Fig. 7. In contrast to the previous case, the energy difference between these two types of states is large, approximately 1 eV. Therefore, the states are clearly divided: $Ga4sp^2$ states are located slightly below the midgap, whereas the $Ga4sp^3$ states are degenerate with the conduction band (CB) and therefore they are empty.

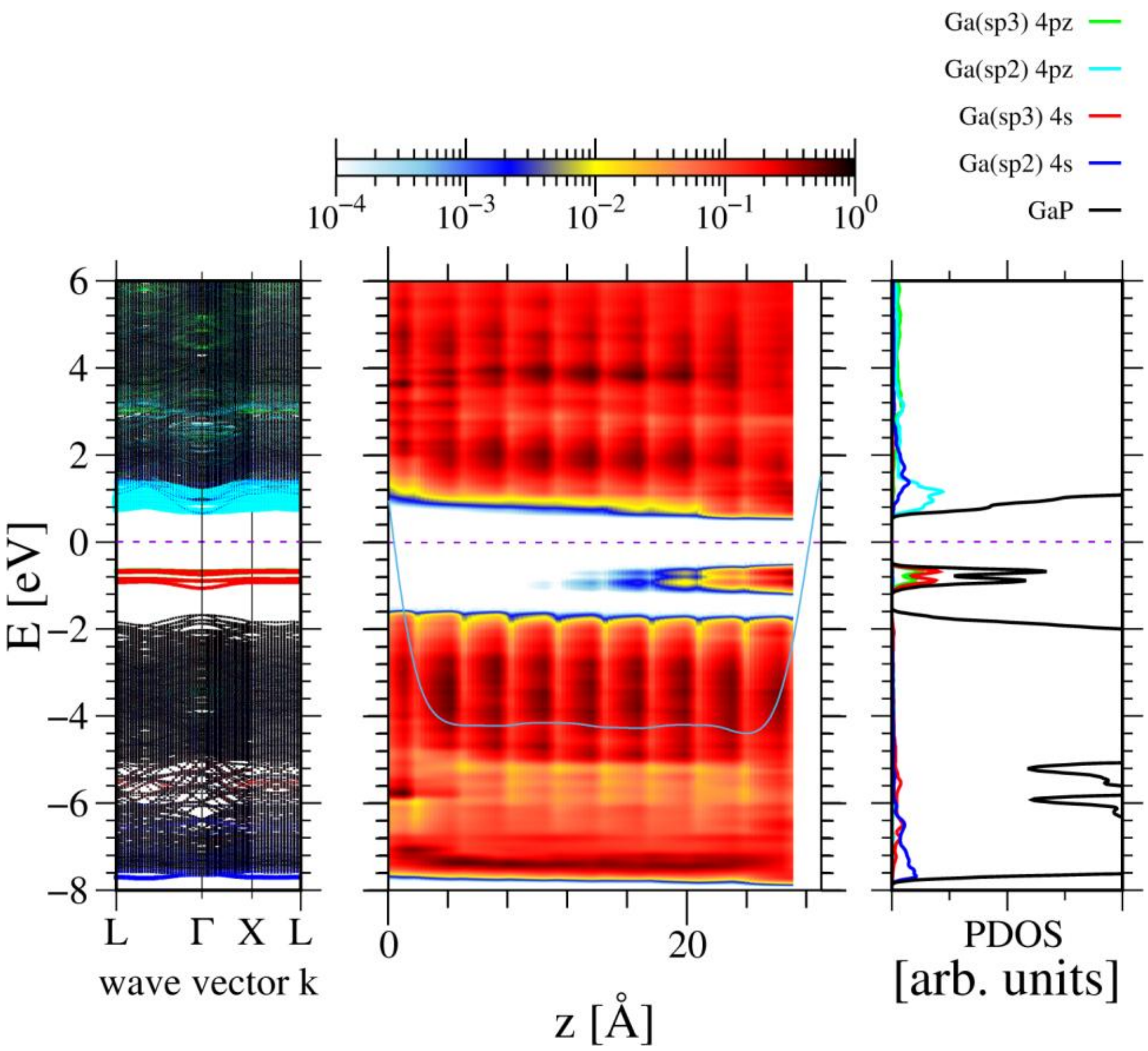

*Fig. 7. Band diagrams in momentum (left panel) and position space (middle panel) of* $(4 \times 4)$ *9 Ga-P DAL slab representing Ga-terminated GaP (111) surface. DOS and PDOS of gallium surface atoms are shown in the right panels.*

These calculations confirmed the predictions based on the Chadi ECR hypothesis, which is also valid for the GaP surface [46]. A charge-based $(4 \times 4)$ reconstruction was observed within the (111) GaP surface. Moreover, the large difference between the energies of these states is reflected in the reconstructed structure in which the difference in the positions of these two sets is large.

The subsequent metal-terminated surface investigated here was a gallium-terminated $GaAs(111)$ surface, represented by the slab plotted in Fig. 8. As shown, a division into $sp^2$ and $sp^3$ hybridized gallium surface atoms is observed. Nevertheless, the number of atoms in the latter configuration is five, which differs from the previously observed $6 \div 10$ divisions.

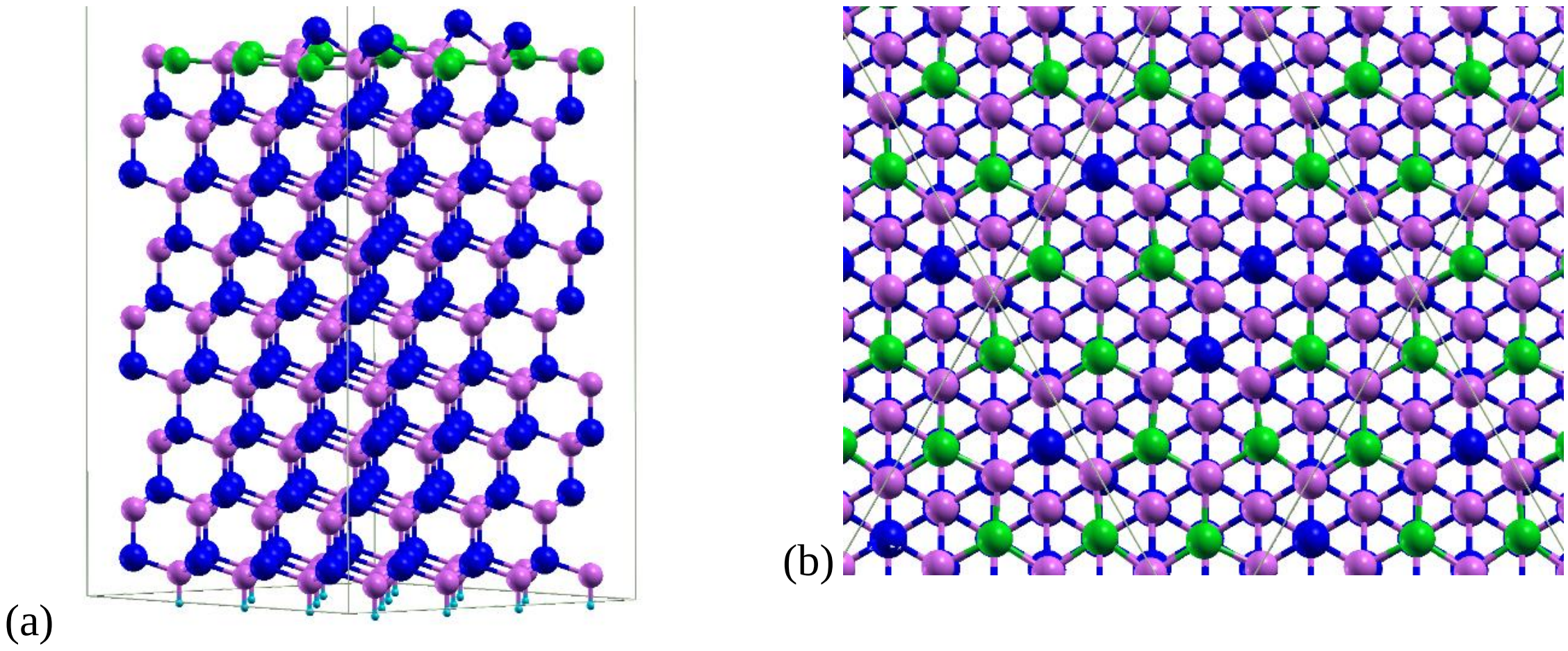

*Fig. 8. Metal-terminated (111)GaAs surface represented by* $(4 \times 4)$ *9 Ga-As DALs slab. Gallium, arsenic and pseudo-hydrogen atoms are denoted by blue, pink and sky-blue balls. The topmost surface atoms in* $sp^2$ *hybridization are denoted by green balls.*

This behavior can be explained by an analysis of the quantum states calculated using $(4 \times 4)$ slab presented in Fig. 9.

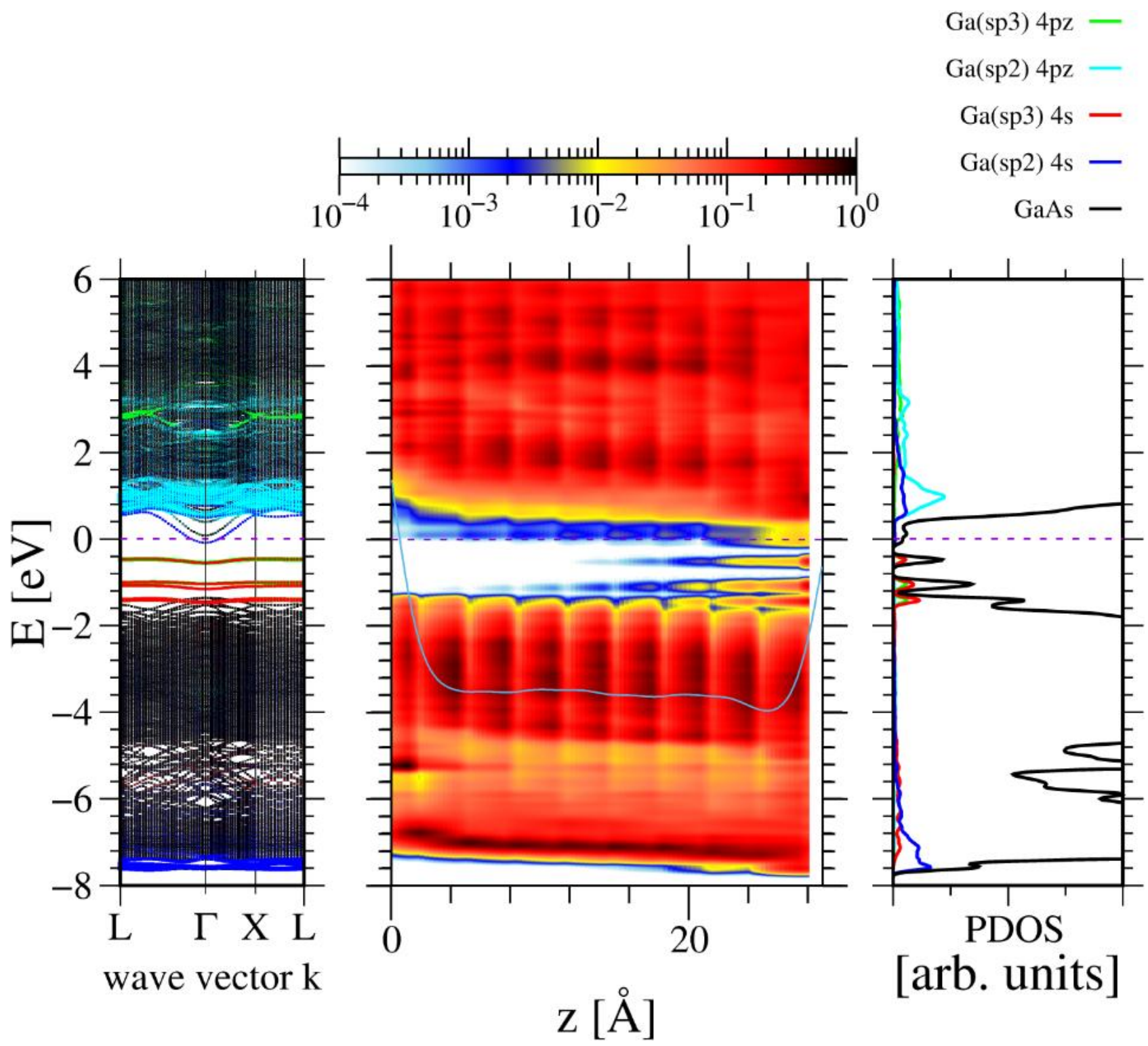

*Fig. 9. Band diagrams in momentum (left panel) and position space (middle panel) of* $(4 \times 4)$ *9 Ga-As DAL slab representing (111) GaAs surface. DOS and PDOS of gallium surface atoms are presented in the right panels.*

The quantum states presented in Fig. 9 are again divided into two sets. The $p_z$ states are degenerate with CB and are therefore empty. The $sp^2$ states are located in the lower portion of the bandgap and are occupied. The Fermi level is within the conduction band; therefore, the electron states are partially occupied. The number of atoms in these two sets is not compatible with the ECR because electrons are partially transferred to the CB. Moreover, the upward positions of these atoms are not as clearly divided as in the case of the GaP surface.

The final case considered is the Ga-terminated $GaSb$ $(111)$ surface, shown in Fig. 9. The division into $sp^2$ and $sp^3$ hybridized sets is again observed, in accordance with the ECR, in $6 \div 10$ proportion.

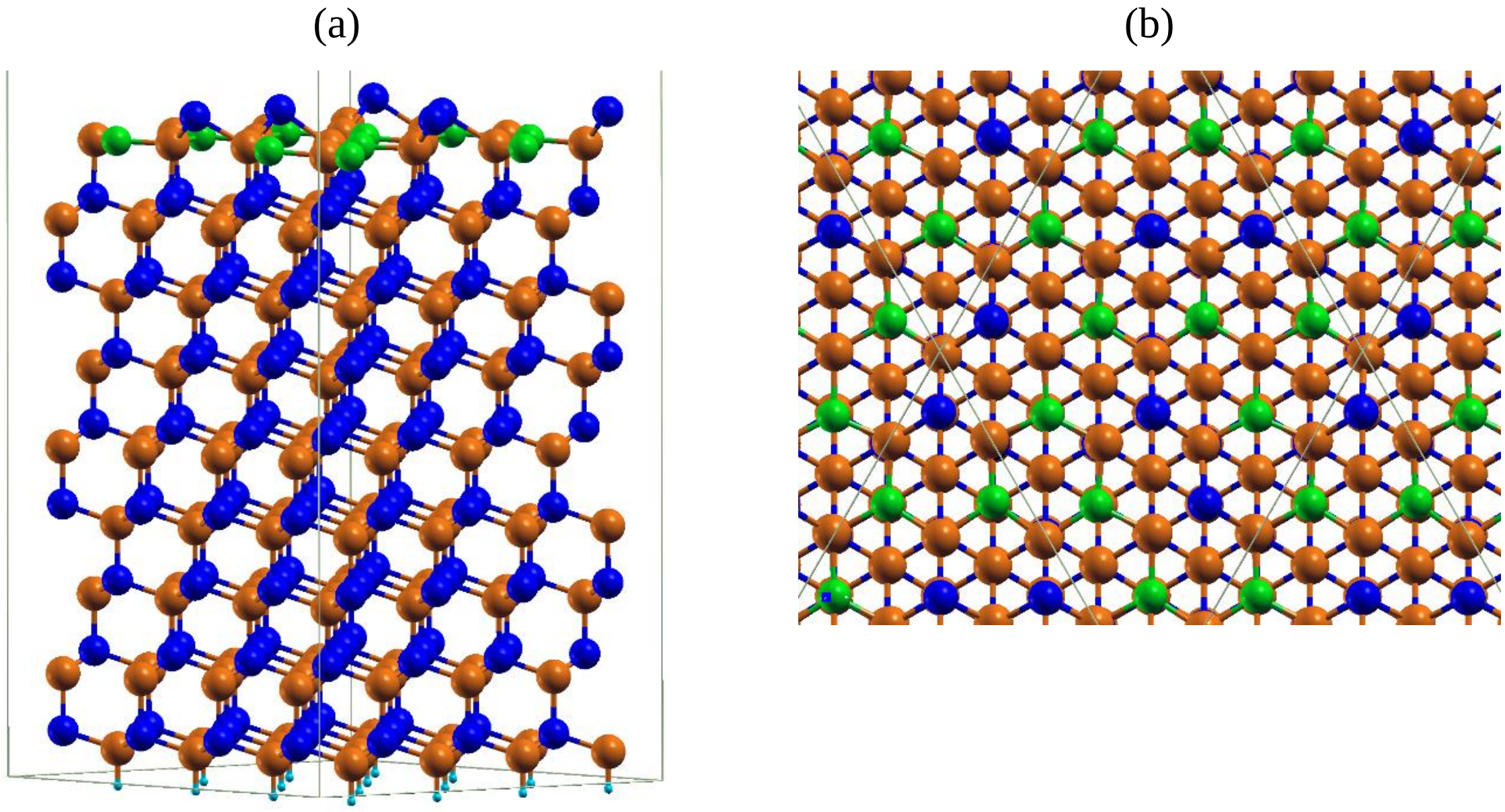

*Fig. 10. Metal-terminated (111) GaSb surface represented by $(4 \times 4)$ 9 Ga-Sb DALs slab. Gallium, antimony and pseudo-hydrogen atoms are denoted by blue, orange and sky-blue balls. The topmost surface atoms in $sp^2$ hybridization are denoted by green balls.*

This conclusion is supported by the slab band diagram shown in Fig. 11. In fact, $sp^2$ states are degenerate with the valence band (VB) and therefore they are fully occupied. The $sp^3$ states are degenerate with the CB and are therefore empty. The bandgap is narrow; therefore, the surface states are located outside the gap, shifted into the bands.

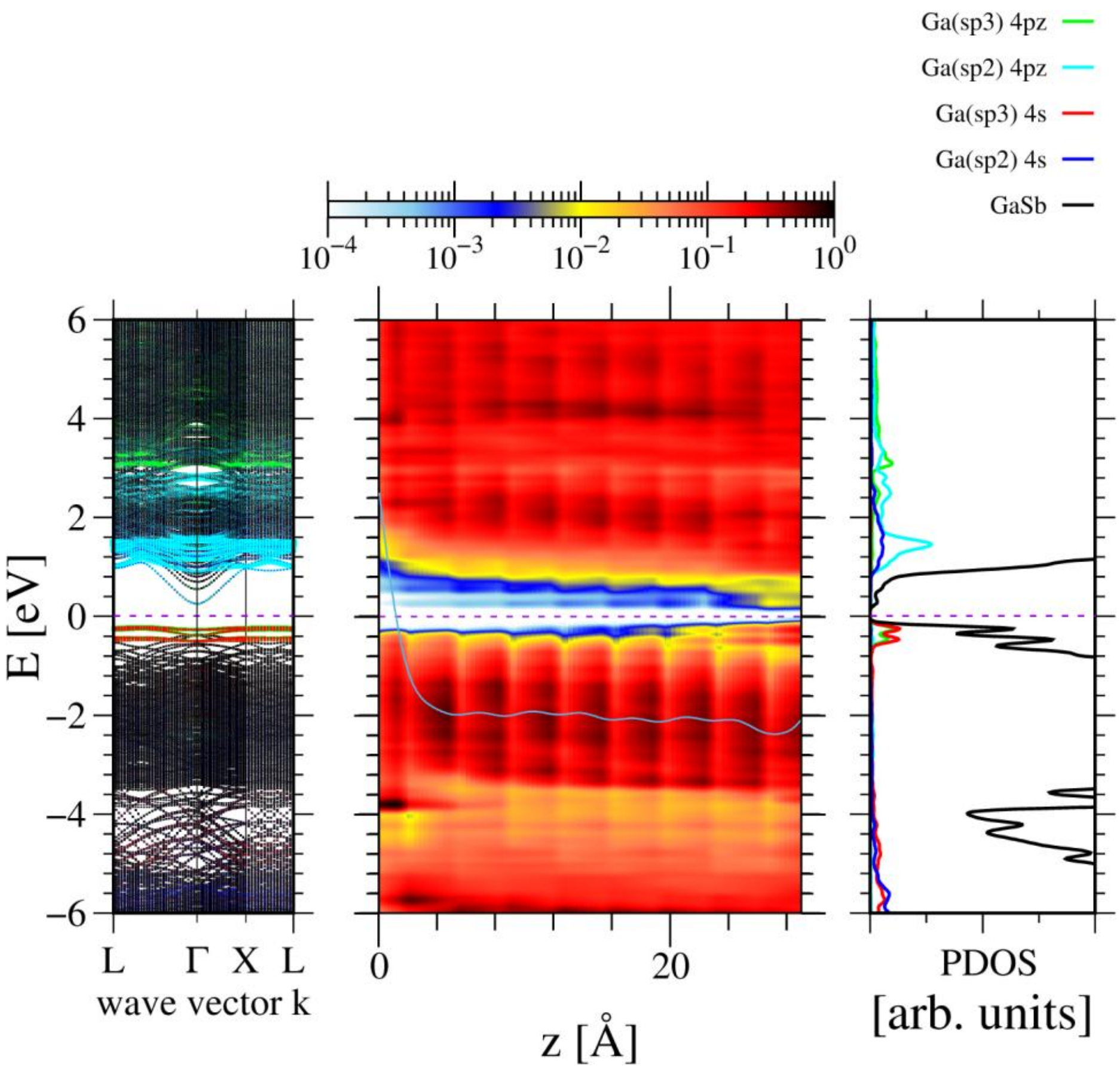

*Fig. 11. Band diagrams in momentum (left panel) and position space (middle panel) of* $(4 \times 4)$ *9 Ga-Sb DAL slab representing (111) GaSb surface. DOS and PDOS of gallium surface atoms are presented in the right panels.*

Table 2. Structural and electronic properties of gallium (Ga)-terminated polar stoichiometric surfaces

| Surface | Structure | CBM - Fermi level difference |
|---|---|---|
| $GaN(0001)$ | $\frac{sp^2}{sp^3} = \frac{3}{5}$ | ~ 0.0 eV |
| $GaP(111)$ | $\frac{sp^2}{sp^3} = \frac{3}{5}$ | 0.8 eV |
| $GaAs(111)$ | $\frac{sp^2}{sp^3} = \frac{5}{11}$ | ~ 0.0 eV |
| $GaSb(111)$ | $\frac{sp^2}{sp^3} = \frac{3}{5}$ | 0.4 eV |

This analysis shows that metal-terminated surfaces also exhibit unified patterns by charge-controlled reconstruction into $(4 \times 4)$ pattern, which is affected by fractional charge exchange with the bulk. Thus, the hypothesis of charge-controlled reconstruction, first proposed in Ref. 43, was confirmed for a wider class of semiconductor surfaces [44].

## IV. Summary and conclusions.

This paper presents an ab initio investigation of the properties of tetrahedral III-V semiconductor polar stoichiometric surfaces. These results are summarized by presenting the state of the art before the present investigation, the results reported in this paper, and the state of the art after completion of this investigation [44].

Following the method presented in Ref. 6, the change of the state of the art is generally presented. The state of the art before the research presented in this work was made, is summarized as follows (6):

(i) The properties of polar metal- and nonmetal-terminated clean surfaces have been investigated only in the case of nitrides.

(ii) The Ga-terminated GaN(0001) surface undergoes ECR charge-controlled $sp^2 - sp^3$ $(4 \times 4)$reconstruction with the Fermi level located between these states.

(iii) Ga broken bond state of $GaN(0001)$ surface is located close to CBM

(iv) N-terminated $(000\bar{1})$ GaN surface is not reconstructed

(v) The energy of the N broken-bond state of $(000\bar{1})$ GaN surface is located close to the VBM, pinning the Fermi level

(vi) The properties of other III-V surfaces have not been determined.

The principal results of this study may be summarized as follows:

(i) Stoichiometric N-terminated $(000\bar{1})$ GaN surface atoms remains in $sp^2$ hybridization

(ii) Nonmetal-terminated surfaces $(\overline{111})$ GaP, $(\overline{111})$ GaAs, $(\bar{1}\bar{1}\bar{1})$ GaSb are not reconstructed, having atoms in $sp^3$ hybridization

(iii) The energies of the nonmetal broken-bond states of $(\bar{1}\bar{1}\bar{1})$ GaP, $(\bar{1}\bar{1}\bar{1})$ GaAs, $(\bar{1}\bar{1}\bar{1})$ GaSb surfaces are located close to the VBM, pinning the Fermi level.

(iv) Ga-terminated surfaces $(111)$ GaP, $(111)$ GaAs, $(111)$ GaSb undergo ECR charge-controlled $sp^2 - sp^3$ $(4 \times 4)$reconstruction,

(v) The Fermi energy is located between $p_z$ and $sp^3$ state energies.

(vi) The valence bands of all compounds are composed of the two separate subbands: the upper composed of $Ga4s$, $Ga4p$ and $NMp$ states and the lower due to overlap of $Ga3d$ and $NMs$ states. Accordingly, these lower energy states are always occupied during surface processes, so they will be omitted in the charge analysis of surface processes.

The resulting state of the art may be summarized as follows:

(i) The principal crystallographic properties of both polar surfaces were determined: nonmetal-terminated surfaces are not reconstructed, and metal-terminated polar surfaces are reconstructed

(ii) The Fermi energy at the metal-terminated polar surfaces is pinned, located high in the bandgap, close to the CBM.

(iii) The Fermi energy at the nonmetal-terminated polar surfaces is pinned by the nonmetal broken-bond state close to the VBM

(iv) The properties of both polar surfaces are drastically different.

(v) The properties of the same-polarity surfaces of chemically different compounds are virtually identical.

In conclusion, the results demonstrate that the determined properties include the crystallographic structure and the electronic properties related to the surface-state energies and occupations. These findings provide a basis for formulating a unified theory of the growth of III-V single crystals and epitaxial layers from the vapor phase based on the similarity of the properties of these surfaces. These investigations will be reported in subsequent papers in this series (II-VII).

**Acknowledgements**

The calculations reported in this paper were performed using the computing facilities of the Interdisciplinary Centre for Mathematical and Computational Modelling of Warsaw University (ICM UW) under grant GB84-23.